\documentstyle{article}
\begin{document}

\title{Kinetics of Conversion of Air Bubbles \\
to Air-Hydrate Crystals in Antarctic Ice}

\author{P. B. Price\\
{\small Physics Department, University of California,
Berkeley, CA 94720}}

\maketitle

\begin{abstract}
The depth-dependence of bubble concentration at pressures above
the transition to the air hydrate phase and the optical scattering length
due to bubbles in deep ice at the South Pole are modeled using
diffusion-growth data from the laboratory, taking into account the
dependence of age and temperature on depth in the ice.  The model
fits the available data on bubbles in cores from Vostok and Byrd and
on scattering length in deep ice at the South Pole.  It explains why
bubbles and air hydrate crystals co-exist in deep ice over a range of
depths as great as 800 m and predicts that at depths below $\rm \sim$ 1400 m
the AMANDA neutrino observatory at the South Pole will operate
unimpaired by light scattering from bubbles.
\end{abstract}

\newpage

Ancient air is known to be trapped in polar ice at depths
below the layer of firn (i.e., porous) ice.  Early investigations
showed that the air was trapped in bubbles which decreased in size
and concentration with increasing depth. To account for the
disappearance of bubbles at great depth, Miller (1) predicted that the
bubbles would convert into a clathrate hydrate phase at depths
corresponding to a pressure greater than that for formation of that
phase. He showed that the phase consists of a cubic crystal structure
in which $\rm O_2$ and $\rm N_2$ molecules from air are trapped in clathrate
cages. If $\rm O_2$ and $\rm N_2$ occur in atmospheric proportions, the
crystals
are usually referred to as air hydrate crystals.  The hatched region in
Fig. 1 shows Miller's calculated curves for the temperature-
dependence of the formation pressure for nitrogen-hydrate and for
air-hydrate, displayed on a scale in which pressure has been
converted to depth in ice.  The curves labeled for ice at four
Antarctic sites and two Greenland sites give temperature as a
function of depth (2-6). Koci (6) modeled the temperature vs depth
at South Pole, using the known surface temperature of -55 C and
fixing the temperature at bedrock at the pressure melting
temperature. {\sl In situ} measurements (7) by AMANDA at depths from
800 to 1000 m gave temperatures that agreed with Koci's model to
within 0.3 C.

In Fig. 1, square points indicate the depths in cores below
which air bubbles are not observed.  The triangular points indicate
the depths below which air hydrate crystals are observed.  Over a
wide range of depths between each triangle and square, both
bubbles and air hydrate crystals are seen to co-exist. For Vostok and
Byrd cores, quantitative measurements have been made of
concentrations and sizes of bubbles (8,9) and air hydrate crystals
(10,11). For Dome C, Dye-3, and Camp Century cores, Shoji and
Langway (12,13) reported only qualitative data on air hydrates. For
the South Pole, no deep core has yet been obtained.

This paper poses solutions to several puzzles:  Why do
bubbles and air hydrate crystals co-exist over a range of depths as
great as 800 m?  In particular, why does it take so long for bubbles
to disappear at pressures at which they are unstable against the phase
transition?  Why do the depths of disappearance of bubbles in
various cores not show some systematic dependence on depth or
temperature?  Based on the measurements on bubbles at Vostok and
Byrd, can we predict the concentration of bubbles as a function of
depth in ice at the South Pole?  The last question is of great
importance to the AMANDA project (14,15), which involves
implanting long strings of large photomultiplier tubes at great depths
in South Pole ice in order to detect \v{C}erenkov light from muons
produced in high-energy neutrino interactions.   Only if the array is
located in bubble-free ice can the direction of a muon be precisely
determined by measuring the arrival times of the \v{C}erenkov
wavefront at each of the phototubes.

Several studies of the transformation of air bubbles into air
hydrate crystals have been done in pressure cells on timescales up to
a few days at temperatures from -20 C to -2 C and at pressures up to
$\rm \sim$ 8 MPa.  Using a high-pressure cell on a microscope stage, Uchida
et al. (16) studied the growth of air hydrate crystals on the walls of
bubbles in a sample of Vostok core taken from a depth of 1514 m.
Because of relaxation after recovery of the core, the original air
hydrate crystals had converted back into bubbles before Uchida et
al. started the experiment.  They observed the growth rate of air
hydrate crystals as a function of supersaturation, $\delta P/P_e$, at
temperatures just below the melting point of ice.  Here $P$ =
hydrostatic pressure on the system and $P_e$ = equilibrium pressure at
the phase boundary.  They found that for $\delta P/P_e > $ 0.35 the crystals
grew as spherical shells coating the bubble walls.

Continuing this line of research, Uchida et al. (17) showed
that two activation energies were involved.  Before a thin shell of air
hydrate crystal had completely coated the wall of a bubble, they
found $E_s$ = 0.52 $\pm$ 0.17 eV.  After it had fully coated the bubble wall,
they found a higher value, $E_s$ = 0.9 $\pm$ 0.1 eV, for thickening of the
shell.  In the early growth stage, it seemed clear that the process
occurred by diffusion of water molecules through the normal ice to
uncoated sites on the bubble wall, because their activation energy
was consistent with that for self-diffusion, 0.57 $\pm$ 0.1 eV (18).  The
higher value, for the later growth stage, applied to diffusion through
the air hydrate itself.  With the assumption (unjustified) of a linear
radial growth rate, they concluded that air hydrate crystals would
form far too quickly to account for the broad range of depths over
which bubbles and air hydrate crystals co-exist in polar ice cores.
They suggested that the rate-limiting process is nucleation, not
diffusion.

Ikeda et al. (19) subjected artificial ice to various hydrostatic
pressures at 270 K and measured the fraction of bubbles converted
to air hydrate crystals in 16 days.  They assumed, without proof,
that the rate-limiting step in the transformation is nucleation.  After
failing to account for the observed rates by homogeneous nucleation
theory, they drew the unsatisfying conclusion that the mechanism
must be heterogeneous nucleation, but with a different parameter for
each data point!  [See their Fig. 7.]

Examination of the various models of nucleation led us to the
conclusion that none of them provides a satisfactory explanation for
the data presented in Fig. 1.  Homogeneous nucleation requires such
an enormous supersaturation, defined as $\delta P/P_e$, that it almost never
occurs in nature.  Heterogeneous nucleation on a foreign surface at
low supersaturation is far more likely.  The presence of a bubble
wall serves as a suitable nucleation site.  Fletcher (20) calculated
nucleation rates as functions of supersaturation, size of the substrate
on which nucleation occurs, and surface energies of the substrate
and the nucleated phase.  For typical bubble sizes and reasonable
values of surface energies, his results show that nucleation would be
rapid at supersaturations below 0.2, whereas for the data in Fig. 1,
bubbles are still present at values of $\delta P/P_e$ as large as 2.  Further,
the presence of one or more screw dislocations in the ice ending at a
bubble surface would reduce the needed supersaturation to a value
less than 0.01 (21).  Typical dislocation densities in even well-
annealed crystals are high enough ($>$ $10^4$ cm$^{-2}$) to ensure their
presence at bubble walls.

I assert that the rate-limiting step in the phase transition is
diffusion rather than nucleation.  To show this I carried out a
diffusion calculation that takes into account the time and temperature
as a function of depth for the Vostok and Byrd sites. I converted
depth to time for each core using age vs depth data in (7). I assumed
that there is no nucleation barrier and that the long time-scale for the
disappearance of bubbles is due to slow diffusion.  I assumed two
diffusion steps.  The first step consists of diffusion of water
molecules through ice to a bubble wall,  in which $D(T)$ is taken to be
$D_0 exp(-E_s/kT)$, with $E_s$ = 0.57 eV and $D_0$ = 1.2 cm$^2$ s$^{-1}$ as
measured (18) for self-diffusion in ice (22).

The second step consists of diffusion of water molecules
through a spherical shell of air hydrate coating the bubble wall and
growing in thickness.   For the activation energy for diffusion in air
hydrate I adopted the value $E_s$ = 0.9 $\pm$ 0.1 eV measured for the
growth of the air hydrate layer (17). [The authors in ref. 17 did not
measure Do.]  After reaching the inner radius of the hollow air
hydrate shell, water combines with air molecules and causes the
crystal to thicken, with negligible activation barrier (23). At
P $\sim$ $10^2$ atm, T $\sim$ 230 K,
the concentration of air in the bubble, $\sim~4 \times 10^{21}$ cm$^{-3}$,
is comparable to that in an air hydrate crystal with the same
volume (8 cages per unit cell; $\rm \sim$ 80\% occupancy; cubic structure;
cube edge = 1.7 nm).  Thus, the supply of air is adequate for full
conversion from (air + ice) to the hydrate phase.

The first step, of water molecules diffusing in ice to the
bubble wall, occurs so rapidly that one can apply the boundary
condition that $C(r > a) = C_0$ for all time, where $a$ is the bubble radius
and $C_0$ is the initial concentration of interstitial $\rm H_{2}O$ molecules
everywhere outside the bubble.  The problem is that of spherically
symmetric diffusion with $C = 0$ at $t = 0$ inside the bubble and $C =
C_0$ at the wall (24).  The justification for taking $C = 0$ at $t = 0$ inside
the bubble is that all of the water vapor inside the bubble ($\rm \sim$
$10^{15}$ molecules cm$^{-3}$ at -40 C) is exhausted in creating an
infinitesimally
thin shell of air hydrate, after which further water must diffuse
through the air hydrate shell (25).

As a function of time, the amount of mass transported
through the bubble wall due to diffusion grows, as given by eq.
6.21 of (24) (with the running index n replaced by j to avoid
ambiguity with my symbol for bubble concentration)

$$ {M(t) \over M(\infty)} = 1 - {6 \over {\pi^2}} \sum_{j=1}^{\infty} {1 \over
j^2} exp(-j^2 \pi^2 D(T)t / a^2) \eqno(1) $$

where $a$ = bubble radius. Equating this to the probability of
disappearance per bubble yields for the fractional concentration
remaining after time $t$

$$ {n(t) \over n_0} = {6 \over {\pi^2}} [exp(-\pi^2 I(t)/a^2) + {1 \over 4}
exp(-4\pi^2 I(t)/a^2) + \cdots ] \eqno(2) $$

$$ {\rm with}~~I(t) = \int_{t_e}^{t} D_0 exp(-E_s/kT) d\acute{t} \eqno(3) $$

The integral $I(t)$ takes into account the fact that $D(T)$ and $t$ change
with depth.  Its lower limit corresponds to the age of the ice at the
transition pressure.  The second term in eq. 2 contributes only at
short times, and higher order terms can be neglected.

I fitted eq. 2 to the extensive Vostok data (8) on bubble
concentrations at depths greater than 500 m, the depth
corresponding to the transition pressure, at which air hydrate
crystals are first observed. With $D_0$ as a fitting parameter, I found
that $D_0$ = 2100 cm$^2$ s$^{-1}$ gave acceptable fits to both the Vostok data
and the Byrd data (9) on bubble concentration as a function of
depth.

Figure 2 displays values of $1/\lambda_{bub} \equiv n{\pi}r^2$, the reciprocal
of
the bubble-to-bubble scattering length, as a function of depth, $z$, for
Vostok, Byrd, and South Pole.  The experimental points use data on
bubble concentration, $n(z)$, and radius, $r(z)$, for Vostok  and for
Byrd.  The data for South Pole are from {\sl in situ} light scattering at
depths of 800 to 1000 m (15).  The curves show the results of
applying the diffusion model to the three sets of data. $n(t)$ is
calculated from eq. 2, taking $a = a_0$, the mean radius at the
dissociation pressure. The observed values are $a_0 = 68 \mu$m for
Vostok and 130 $\mu$m for Byrd. In the absence of data on $a_0$ for the
South Pole, I assumed the same value as for Vostok, since those
two sites have similar elevations, surface temperatures, atmospheric
pressures, and hydrate dissociation pressures (see Fig. 1), which
are rather different from those at Byrd. To compute the curve for
$1/\lambda_{bub}$, I assumed that $r = a_0 (P_e/P)^{1 \over 3}$
due to hydrostatic pressure.
The fits to the data for Vostok and Byrd are quite good and lend
confidence to the predicted dependence of $1/\lambda_{bub}$ on depth for the
South Pole (26).

The diffusion-growth model provides a solution to the
puzzles listed in the introduction.  The reason that bubbles do not all
convert into air hydrate crystals at the phase transition pressure, and
the reason for the great range of depths at which both air hydrate
crystals and bubbles co-exist, is that the time required for water
molecules to diffuse through a growing shell of air hydrate at
ambient ice temperature is extremely long.  The diffusion coefficient
for water in air hydrate, $D(T)$ = $D_0 exp(-E_s/kT)$, with $D_0$ = 2100 cm$^2$
s$^{-1}$, is orders of magnitude smaller than for self-diffusion in
hexagonal ice. For example, at -46 C, the temperature of South Pole
ice at a depth of 1 km, $D$ = 2.2 $\times$ $10^{-17}$ cm$^2$ s$^{-1}$
for water in air
hydrate, whereas $D$ = 2.65 $\times$ $10^{-13}$ cm$^2$ s$^{-1}$
for water in hexagonal
ice. The reason for the apparent lack of organization of the data on
bubble disappearance in Fig. 1 is that the depth is the wrong variable
to use.  Due to large variations in snow accumulation rate from one
polar site to another, depth is not universally related to time.  Only
when data are plotted on a graph of time vs reciprocal of temperature
does the correlation become clear.  When applied to laboratory data
on rate of decrease of concentration of bubbles in ice near the
melting point (19), the model gives results consistent with the data.

The model predicts that in deep ice at the South Pole, the
bubble-to-bubble scattering length is $\rm \sim$ 6 m at 1300 m, 20 m at 1400
m, and 130 m at 1500 m.  For an AMANDA phototube spacing of
20 m, bubbles will cease to degrade imaging at depths greater than
$\rm \sim$ 1400 m.

This work was supported in part by National Science
Foundation grant PHY-9307420.

\newpage

\subsection*{REFERENCES}

\begin{list}{}{}
\item[1.]{S.L. Miller, {\sl Science} {\bf 165}, 489 (1969).}
\item[2.]{For Dome C, see C. Ritz, L. Lliboutry, C. Rado, {\sl Ann.
Glaciology} {\bf 3}, 284 (1982).}
\item[3.]{For Dye 3, see D. Dahl-Jensen and S. J. Johnsen, {\sl Nature} {\bf
320},
250 (1986).}
\item[4.]{For Byrd and Camp Century, see W. F. Budd and N. W.
Young, in {\sl The Climatic Record in Polar Ice Sheets}, G. de Q. Robin,
Ed. (Cambridge Univ. Press, Cambridge, 1983), p. 150.}
\item[5.]{For Vostok, see C. Ritz, {\sl Ann. Glaciology} {\bf 12}, 138 (1989).}
\item[6.]{A model for temperature vs depth at the South Pole was
developed by Bruce Koci, unpublished material.}
\item[7.]{AMANDA Collaboration (P. Askebjer {\sl et al.}),
submitted to {\sl J. Glaciology} (1994).}
\item[8.]{N.I. Barkov and V.Ya. Lipenkov, {\sl Mat. Glyatsiol. Issled.}
{\bf 51}, 178 (1984).}
\item[9.]{A.J. Gow and T. Williamson, {\sl J. Geophys. Res.} {\bf 80},
5101 (1975).}
\item[10.]{V.Ya. Lipenkov, {\sl Mat. Glyatsiol. Issled.} {\bf 65}, 58
(1989).}
\item[11.]{T. Uchida, T. Hondoh, S. Mae, V. Ya. Lipenkov, P. Duval,
{\sl J. Glaciology} {\bf 40}, 79 (1994).}
\item[12.]{H. Shoji and C.C. Langway, Jr., {\sl Nature} {\bf 298}, 548
(1982).}
\item[13.]{H. Shoji and C.C. Langway, Jr., {\sl J. Phys. (Paris)} {\bf
48}, colloq. C1, 551 (Supplement au 3) (1987).}
\item[14.]{S.W. Barwick et al., {\sl J. Phys. G (Nucl. Part. Phys.)}
{\bf 18}, 225 (1992).}
\item[15.]{AMANDA Collaboration (P. Askebjer {\sl et al.}),
submitted to {\sl Science} (1994).}
\item[16.]{T. Uchida et al., in {\sl Proc. Inter. Conf. on Physics and
Chemistry of Ice} (ed. N. Maeno and T. Hondoh), Hokkaido University Press,
Sapporo, 1992, pp. 121-125.}
\item[17.]{T. Uchida et al., in {\sl Fifth Inter. Symp. On Antarctic
Glaciology}, to be published (1994).}
\item[18.]{K. Goto, T. Hondoh, A. Higashi, {\sl Jap. J. Appl. Phys.} {\bf
25}, 351 (1986).}
\item[19.]{T. Ikeda, T. Uchida, S. Mae, {\sl Proc. NIPR Symp. Polar
Meteorog. Glaciol.} {\bf 7}, 14 (1993).}
\item[20.]{N. H. Fletcher, {\sl J. Chem. Phys.} {\bf 29}, 572 (1958).}
\item[21.]{W.K. Burton, N. Cabrera, F.C. Frank, {\sl Phil. Trans. Roy.
Soc. (London)} {\bf A 243}, 299 (1951).}
\item[22.]{ The authors of (18) showed that self-diffusion in ice occurs by
diffusion of interstitial water molecules rather than vacancies.}
\item[23.]{An alternative possibility, that air molecules diffuse through the
air hydrate shell and convert ice to air hydrate at the bubble
boundary, is ruled out experimentally.  Uchida et al. (16,17)
observed that air hydrate crystals grow inwardly, causing bubbles to
shrink in size and eventually disappear.}
\item[24.]{J. Crank, {\sl The Mathematics of Diffusion} (Clarendon Press,
Oxford, 1975).}
\item[25.]{This assumes that the vapor pressure of water at the surface of
an air hydrate crystal is much smaller than at the surface of an ice
crystal.}
\item[26.]{An alternative procedure is to fix $D_0$ = 1.2 cm$^2$ s$^{-1}$,
the same for
diffusion of water in air hydrate as in ice, and to treat $E_s$ as a fitting
parameter.  Doing this leads to acceptable fits to the Vostok and
Byrd data but with $E_s$ = 0.75 eV, which is outside the standard error
claimed in (11).  The resulting curves for $1/\lambda_{bub}$ for Vostok, Byrd,
and South Pole are  similar to those calculated with $D_0$ = 2100 cm$^2$
s$^{-1}$, $E_s$ = 0.9 eV.  The conclusion is that the present data do not
enable one to determine both $D_0$ and $E_s$ independent of laboratory
data.}
\end{list}

\newpage

\subsection*{FIGURE CAPTIONS}

\newcounter{capnum}

\begin{list}{\arabic{capnum}.}{\usecounter{capnum}}
\item Temperatures as a function of depth in ice, compared with
equilibrium pressures (converted to depths) for co-existence of the
(bubble + ice) phase and the air hydrate phase.  Upper boundary of
the hatched region is for nitrogen-clathrate-hydrate; lower boundary
is for air-clathrate-hydrate. Square points indicate the depths in cores
below which air bubbles are not observed. The arrow on the square
symbol for Dome C indicates that air hydrate crystals were present
to the bottom of the core. Triangular points indicate the depths
below which air hydrate crystals are observed.
\item Reciprocal of bubble-to-bubble scattering length, $1/\lambda_{bub} =
n{\pi}r^2$, as function of depth.  Data for Vostok (8) and Byrd (9) were
obtained with microscopic examination of core samples in a cold
laboratory; data for South Pole were obtained by {\sl in situ}
measurement of optical scattering (15).  Upper solid triangles
assume forward scattering from smooth-walled bubbles; lower solid
triangles assume isotropic scattering from rough-walled bubbles.
Hydrostatic pressure curve shows effect of shrinkage of bubble size
without change of concentration.  Calculated curves for the three
sites show the effect of a decrease in bubble concentration as a
function of time due to diffusion of $\rm H_{2}O$ molecules through air
hydrate crystal walls.
\end{list}

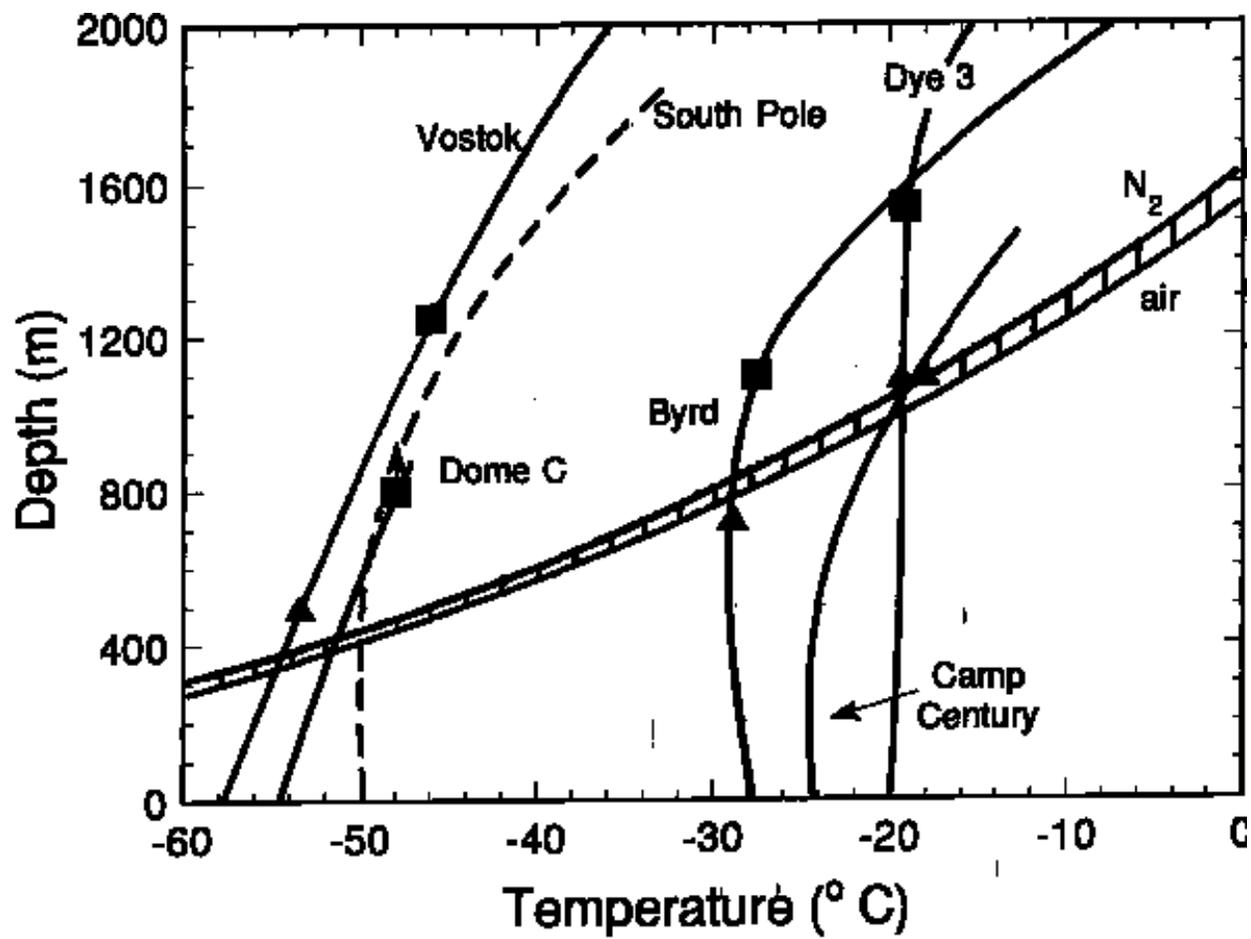
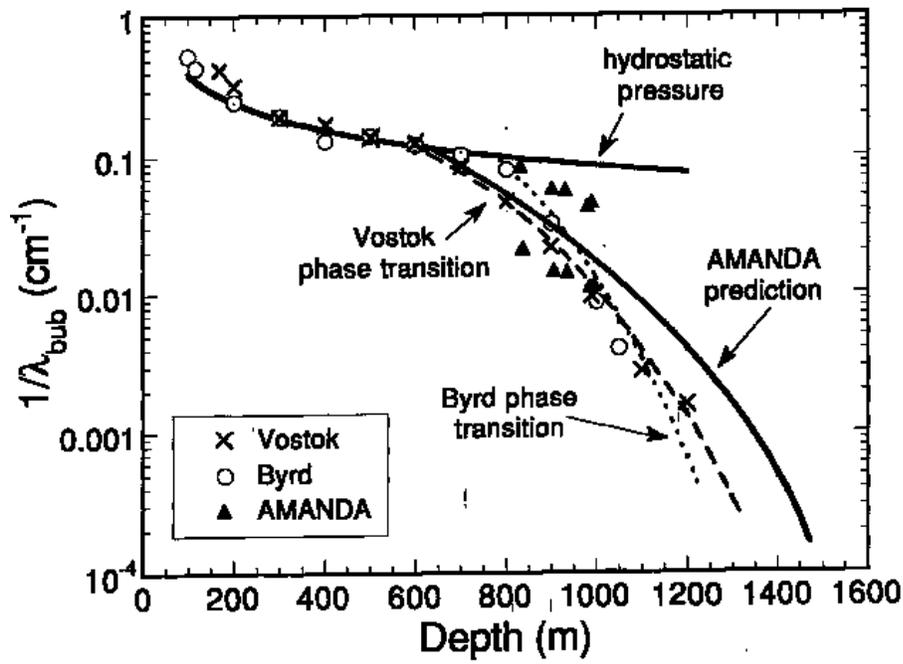
\end{document}